\documentclass[amsmath,amssymb,showpacs,floatfix,aps,prl,twocolumn,superscriptaddress]{revtex4-1}
\usepackage{graphics}
\usepackage{dcolumn}

\begin{document}


\title{Semiconducting Monolayer Materials as a Tunable Platform for Excitonic Solar Cells}
\author{Marco Bernardi}
\affiliation{%
Department of Materials Science and Engineering, Massachusetts Institute of Technology, \\
77 Massachusetts Avenue, Cambridge MA 02139-4307, USA}
\author{Maurizia Palummo}
\affiliation{%
Department of Materials Science and Engineering, Massachusetts Institute of Technology, \\
77 Massachusetts Avenue, Cambridge MA 02139-4307, USA}
\affiliation{%
Dipartimento di Fisica, Universit$\grave{a}$ di Roma Tor Vergata, CNISM, and European Theoretical Spectroscopy Facility (ETSF), Via della Ricerca Scientifica 1, 00133 Roma, Italy}

\author{Jeffrey C. Grossman}
\email[E-mail: ]{jcg@mit.edu}
\affiliation{%
Department of Materials Science and Engineering, Massachusetts Institute of Technology, \\
77 Massachusetts Avenue, Cambridge MA 02139-4307, USA}

\date{\today}

\begin{abstract}
The recent advent of two-dimensional monolayer materials with tunable optoelectronic properties and high carrier mobility offers renewed opportunities for efficient, ultra-thin excitonic solar cells alternative to those based on conjugated polymer and small molecule donors. Using first-principles density functional theory and many-body calculations, we demonstrate that monolayers of hexagonal BN and graphene (CBN) combined with commonly used acceptors such as PCBM fullerene or semiconducting carbon nanotubes can provide excitonic solar cells with tunable absorber gap, donor-acceptor interface band alignment, and power conversion efficiency, as well as novel device architectures. For the case of CBN-PCBM devices, we predict the limit of power conversion efficiencies to be in the $10-20$\% range depending on the CBN monolayer structure. Our results demonstrate the possibility of using monolayer materials in tunable, efficient, polymer-free thin-film solar cells in which unexplored exciton and carrier transport regimes are at play.
\end{abstract}


\maketitle

\section*{}
Solar cell devices converting energy radiated from the sun to electricity have developed into two main families: those based on bulk inorganic semiconductors such as Si, GaAs, CdTe, and CIGS \cite{mrsreview}, in which free charge carrier generation follows light absorption without intermediate steps,
and those based on conjugated polymers and small molecules \cite{mcgeheereview, nelsonreview} or other materials where following light absorption a complex of hole and electron carriers (exciton) is formed with a binding energy in large excess of $kT$. 
The latter type, referred to as excitonic solar cell (XSC)\cite{excpv1}, realizes charge carrier generation by dissociating bound excitons at semiconductor heterointerfaces, owing to discontinuities across the interface in the electron affinity and ionization potential.\\
\indent
A typical solid state XSC \footnote{
Our discussion here does not include dye-sensitized solar cells (DSSC), in which due to the presence of a liquid phase the carrier and exciton dynamics are different than in the solid state. For a recent review of DSSC, see M. K. Nazeeruddin \textit{et al.}, Solar Energy 85 (2011) 1172.} employs a donor-acceptor blend of conjugated polymer or small molecule donors with high optical absorption in the visible, and fullerene derivative acceptors (\textit{e.g.} $\mathrm{C_{60}}$, PCBM or similar fullerene molecules) \cite{mcgeheereview, pc70bm, adducts}. In such devices, the polymer donor dominates (and limits) the key physical processes in the solar cell, including optical absorption and transport of excitons and charge carriers. In particular, exciton diffusion lengths of $5-10$ nm prevent the use of bilayer devices, and mobilities lower than 1 $\mathrm{cm^2 / {Vs}}$ limit the thickness of bulk heterojunction devices to less than the absorption depth (typically $0.1-1\mu$m).\\ 
\indent
In addition, tuning the HOMO and LUMO levels of conjugated polymers is a challenging task, requiring trial and error chemical synthesis of a large number of compounds; band gaps of less than 1.5 eV are hard to achieve, and thus the absorption loss in the red part of the solar spectrum can be significant \cite{nelsonreview}. Carrier transport in conjugated molecules occurs via polarons in a regime of strong electron-phonon coupling, leading to ultrafast photoexcited carrier relaxation and consequent thermalization loss. 
Despite such inherent material limitations and the related constraints they place on the device architecture, polymer and small molecule XSC technologies have progressed to impressive power conversion efficiencies, currently up to approximately 11\% \cite{Heliatex, UCLA}.\\  
\indent
Alternative XSC technologies have emerged in recent years, including efficient quantum dot based \cite{sargent} and more recently nanocarbon based XSC \cite{acpvpreprint}.
The key advantage of these novel excitonic devices is the possibility of altering the HOMO and LUMO levels, the band alignment and the optical absorption by using quantum confinement in nanomaterials rather than different chemistries as in the case of polymers and small molecules. Of great relevance is also the possibility of novel device architectures \cite{sargent}, while retaining the inherent advantages of solution-based, low temperature manufacturing on flexible substrates typical of XSC.\\
\indent
In this context, the recent advent of two-dimensional (2D) monolayer materials with tunable optoelectronic properties and high carrier mobility $-$ \textit{e.g.} monolayers of graphene \cite{geim}, graphene-BN \cite{Aja, ourcbn}, $\mathrm{MoS_2}$ \cite{mos2photoluminescence, mos2, Mos2exc}, and other transition-metal oxides and dichalcogenides \cite{Can} $-$ offers renewed opportunities for thin-film XSC. Although graphene is a semimetal that can only be used in Schottky-type solar cells \cite{nicola}, other monolayers are semiconductors with strong excitonic effects and form excellent candidates for XSC. Due to the peculiar nature of transport and electron-phonon coupling in monolayer materials \cite{levitov, pablo, Columbia}, novel regimes for photovoltaic operation can be envisioned, such as hot carrier extraction, multiple exciton generation and coherent exciton transport. 
Even ultra-thin devices may effectively capture sunlight due to the high optical absorption of 2D monolayers: for example, one layer of graphene absorbs 2.3\% of the incident intensity in the visible \cite{nair}, with a van der Waals stacking thickness of only 3.3 \AA \hspace{2pt}in the direction normal to the layer.
Tunable optoelectronic properties can be achieved by either controlling the structure and quantum confinement within the layer, or by stacking different monolayers to create novel van der Waals structures. An example of the former strategy is the case of hybridized graphene-BN (CBN) monolayers \cite{Aja}, whose electronic band gap, optical absorption and exciton binding energy can be varied by tuning the C domain size and shape, due to quantum confinement of excitons within the C domains. 
The mobility in such CBN layers synthesized by CVD can be as high as 20 $\mathrm{cm^2 / Vs}$ \cite{Aja, ourcbn}.\\
\begin{figure}
\includegraphics{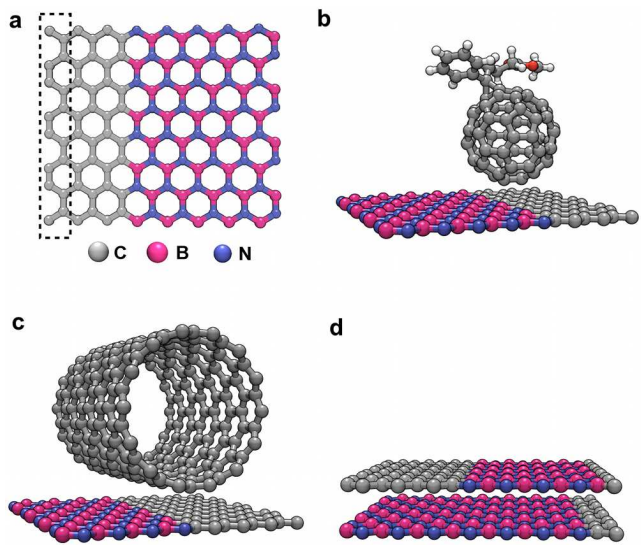}
\caption{(a) CBN monolayer unit cell used in the DFT calculations, with an armchair edge between the C and BN domains. Shown in the dashed box is a C atom row, with an approximate width of 0.25 nm. Following the nomenclature used here, the unit cell shown is $\mathrm{C_3(BN)_5}$. Panels (b) to (d) show interfaces between a CBN monolayer and (b) PCBM fullerene, (c) a (14,0) zig-zag SWCNT, and (d) another CBN layer with different composition. \label{fig1}}
\end{figure}
\indent

Here we show that interfaces between CBN monolayers and carbon-based acceptors such as PCBM and semiconducting single-walled carbon nanotubes (s-SWCNT) can form tunable type-II band alignments \cite{franceschetti}, and are thus suitable to realize exciton dissociation in 2D monolayer based XSC. The power conversion efficiency limit resulting from such CBN interfaces is shown to be tunable, and is estimated to be in the $10-20$\% range. Our calculations further suggest that even just two stacked monolayers of CBN 
with the proper structure could constitute an ultra-thin solar cell with $\approx$ 3 \AA \hspace{2pt}thickness. We propose device architectures for the experimental study of transport, excited state dynamics and power conversion efficiency in XSC based on monolayer materials.\\ 
\indent
We employ \textit{ab initio} density functional theory (DFT) calculations using the QUANTUM ESPRESSO code \cite{QE} on model CBN systems constituted by a monolayer with C and BN stripe domains arranged in a 2D superlattice and separated by an armchair edge. The CBN repeat unit consists of a layer with 8 atom rows and an overall composition of $\mathrm{C_x(BN)_{(8-x)}}$, where $x$ is the number of C rows in the structure, each 0.25 nm wide (Fig. \ref{fig1}(a)) \footnote{
Though other geometries for C and BN domains are possible, such as quantum dots of BN in C (or viceversa), the results presented here refer to structures lacking confinement in one of the two directions within the monolayer. When confinement is introduced in both directions by forming dots (\textit{e.g.} see J. Li and V. B. Shenoy, Appl. Phys. Lett. \textbf{98}, 013105, 2011), the DFT band gaps are usually higher than those found here.}.\\
\indent
Interfaces containing CBN sheets (Fig. \ref{fig1}(b-d)) are formed by placing PCBM or s-SWCNT of chirality (10,0), (14,0), and (16,0), or a CBN layer at a Van der Waals distance of 3.3 \AA \hspace{2pt} from the CBN monolayer. An orthorhombic simulation cell was adopted, and all structures are fully relaxed within DFT to less than 30 meV/\AA  \hspace{2pt}in the residual atomic forces. A 15 \AA  \hspace{2pt} vacuum is placed in the direction normal to the sheet to avoid spurious interactions with the image system. The Perdew-Burke-Ernzerhof exchange-correlation functional \cite{PBE} is adopted and ultrasoft psuedopotentials \cite{USPP} are used to describe the core electrons. A kinetic energy cutoff of 35 Ry was used for the plane-wave basis set and of 200 Ry for the charge density, in combination with converged Monkhorst-Pack $\vec{k}$-point grids \cite{Kgrid} of up to $24\times8\times1$. For the CBN bilayer calculations, the unit cells consisted of two $\mathrm{C_x(BN)_{(8-x)}}$ layers with AB stacking in the BN domains (B atoms on top of N atoms) \cite{bnlayers}, and the vdW-DF exchange-correlation functional \cite{vdwlayered} was employed as implemented in QUANTUM ESPRESSO.\\ 
\indent
For all the interfaces studied here, the DFT derived HOMO and LUMO level offsets were estimated as differences in the peaks of the projected density of states (PDOS) for the two structures constituting the interface, similar to Ref. \cite{yosuke}. Though this method is usually reliable to estimate valence band offsets \cite{gianmarco}, accurate conduction band offsets can only be obtained if the DFT error on the band gap is comparable for the two materials constituting the interface. In order to estimate the conduction band offsets with higher accuracy, for the CBN-PCBM interfaces we applied $GW$ corrections \cite{gianmarco} separately to the band gaps of both materials, and the $GW$+Bethe-Salpeter equation ($GW$+BSE) formalism to compute the optical gap of the CBN monolayers. For the CBN layers, both the $GW$ corrections and the $GW$+BSE optical gaps for three compositions ($\mathrm{C_1(BN)_7}$, $\mathrm{C_4(BN)_4}$ and $\mathrm{C_7(BN)_1}$) were taken from our previous work \cite{ourcbn}. The corrections decrease linearly for increasing C domain size, so that corrections at intermediate compositions were derived by interpolation. The $GW$ correction for the PCBM electronic gap was calculated here using the Yambo code \cite{Yambo} within a $G_0W_0$ update scheme, using a plasmon pole model for the dielectric function together with cutoffs of 35 Ry and 8 Ry for the exchange and correlation part of the self-energy, and up to 1500 empty bands. Further details of the many-body calculations are reported in the Supporting Material.
\begin{figure*}[!ht]
\includegraphics{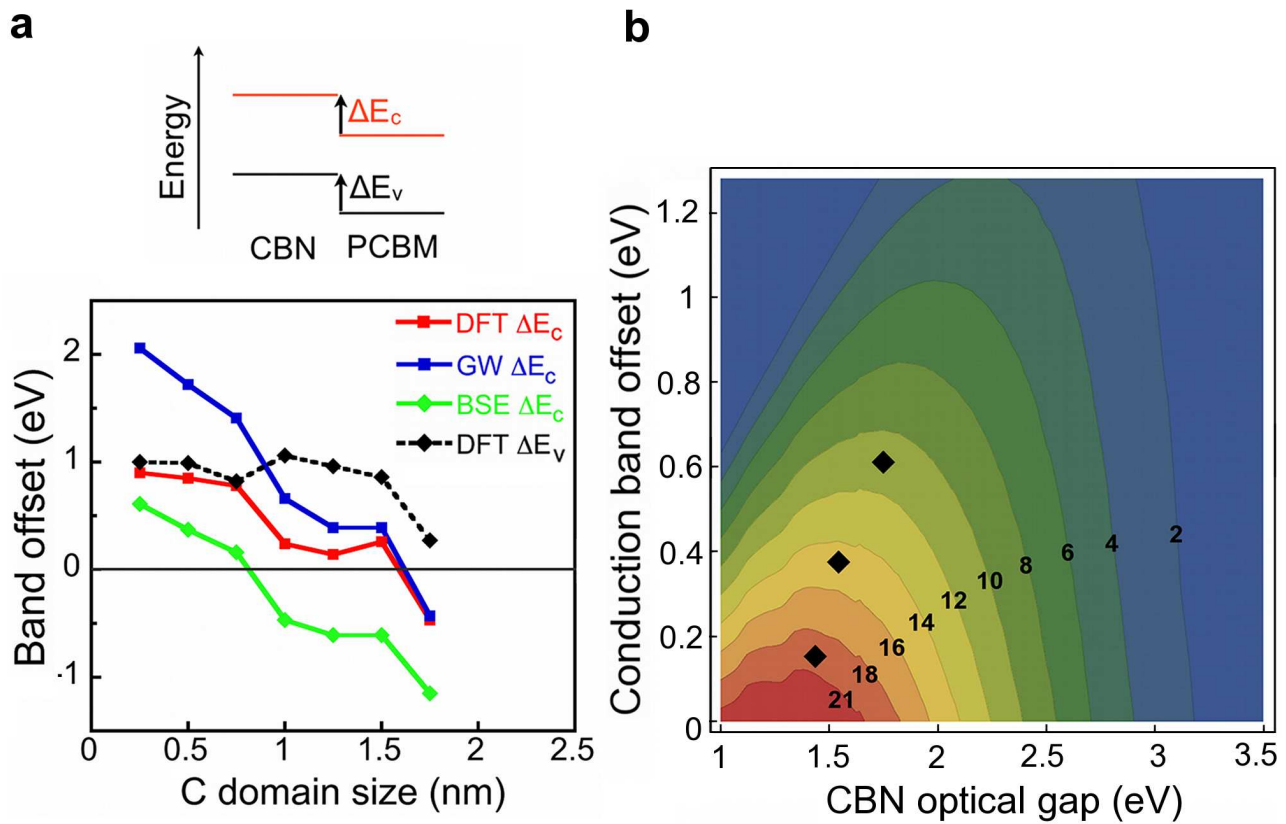}
\caption{(a) DFT valence band offset ($\Delta E_v$, dashed curve) and DFT, $GW$ and BSE conduction band offsets ($\Delta E_c$, solid curves) between $\mathrm{C_x(BN)_{(8-x)}}$ monolayers and PCBM fullerene, expressed as a function of the C domain size in the CBN layer. $\Delta E_v$ and $\Delta E_c$ are referenced, respectively, to the valence and conduction band edges of the acceptor, as shown schematically above the plot. Within our model, $\Delta E_v$ is the same at all levels of theory, since the  corrections fall entirely on the conduction band. The number $x$ of C atom rows in the CBN superlattice structure increases by one unit (from 1 to 7) at each consecutive point in the plot, and for each added row the C lateral domain size increases by 0.25 nm. The alignment is consistently found to be type-II for C domain sizes of up to approximately $1-2$ nm depending on the level of theory used. (b) Power conversion efficiency contour plot as a function of the CBN donor optical gap and conduction band offset $\Delta E_c$. Constant efficiency level curves up to 21\% are shown in figure. The diamonds represent the efficiency limits of the three CBN-PCBM combinations yielding type-II alignment in the BSE curve in (a). Efficiency values in the $10-20$\% range are predicted depending on the C domain size. \label{fig2}}
\end{figure*}
\indent
Fig. \ref{fig2}(a) shows the valence and conduction band offsets ($\Delta E_v$ and $\Delta E_c$, respectively) at different levels of theory for interfaces between PCBM and $\mathrm{C_x(BN)_{(8-x)}}$ monolayers with different C domain sizes. Each consecutive point represents the addition of one C atom row in the unit cell (Fig. \ref{fig1}(a)), leading to a 0.25 nm increase of the C domain size. The DFT band alignment is found to be type-II for C domain sizes of up to $1.5$ nm, with the PCBM acting as the acceptor at the interface, as seen by the positive values of $\Delta E_v$ and $\Delta E_c$ using the convention shown in the figure. The trends in $\Delta E_v$ and $\Delta E_c$ show that the HOMO and LUMO energies of the donor can be tuned according to the C domain size, yielding unique control over the interface band offsets. This results in a tunable power conversion efficiency, as explained below.\\ 
\indent
To confirm the type-II alignment found within DFT, we apply the $GW$ correction separately to the band gaps of CBN and PCBM (see Supporting Material). The $GW$ corrected $\Delta E_c$ values \footnote{
Note that in this work $\Delta E_v$ is the same at all levels of theory, since the $GW$ corrections found here fall almost entirely on the LUMO level and were thus applied as a scissor correction, as further explained in the Supporting Material.
}, shown in Fig. \ref{fig2}(a) ($GW$ curve) yield the same qualitative trends as the DFT results, with type-II alignment for C domain sizes of up to 1.5 nm.\\ 
\indent
Assuming that the CBN layer donor is also the main absorber in the XSC, the exciton binding energy in the CBN layer is the key quantity to determine the possibility to transfer a photoexcited electron to the PCBM. To address this point, we derived  $\Delta E_c$ values as differences between the LUMO level derived from the optical gap of the CBN donor (using $GW$+BSE) and the $GW$ LUMO level of the PCBM acceptor. This \textit{combined scheme} utilizing the optical LUMO of the donor and the quasiparticle LUMO of the acceptor can correctly describe the minimum energy of the exciton formed after photoabsorption in the CBN donor, as well as the correct electronic quasiparticle level for the addition of the transferred electron to the acceptor. The $\Delta E_c$ values derived at this combined level of theory are shown in Fig. \ref{fig2}(a) (BSE  $\Delta E_c$ curve), and are used below to calculate the power conversion efficiencies; within this approximation, the useful range for XSC operation is restricted to C domain sizes of up to approximately 1 nm, for which the $\Delta E_c > 0$ condition is met
\footnote{Given the choice of working with superlattices instead of dots geometries, we remark that this is a lower bound of the C domain size for this effect to be observed in practice.}.\\ 
\indent
For the three CBN-PCBM cases satisfying this condition, we estimate a practical limit to the power conversion efficiency. The thermodynamic limit to the efficiency for thermalized carriers and in the absence of non-radiative recombination is set by the optical gap of the donor through the Schockley-Quisser limit \cite{SQ}. However, efficiency \textit{trends and practical limits} are far more useful for XSC than ultimate thermodynamic limits \cite{scharber, lunt}.\\
\indent
Following Scharber \textit{et al.} \cite{scharber}, we estimate practical values of the maximum power conversion efficiency $\eta$ for CBN-PCBM devices with type-II alignment as:
\begin{equation}
\label{eq1}
\eta = \frac{0.65 \cdot (E_{g}^{\mathrm{opt, d}} - \Delta E_c - 0.3) \cdot \int_{E_{g}^{\mathrm{opt, d}}}^{\infty} \frac{J_{ph}(\hbar \omega)}{\hbar \omega} \,d(\hbar \omega)}{\int_{E_{g}^{\mathrm{opt, d}}}^{\infty} J_{ph}(\hbar \omega) \,d(\hbar \omega)}
\end{equation}

where 0.65 is the fill factor ($FF$), $J_{ph}(\hbar \omega)$ is the AM1.5 solar energy flux (expressed in $\mathrm{W\,m^{-2}\,eV^{-1}}$) \cite{NREL} at the photon energy $\hbar \omega$, and $E_{g}^{\mathrm{opt, d}}$ is the optical band gap of the CBN donor.\\ 
\indent
In Eq. \eqref{eq1}, the $(E_{g}^{\mathrm{opt, d}} - \Delta E_c - 0.3)$ term is an estimate of the maximum open circuit voltage ($V_{oc}$, in eV), calculated as the effective interface gap $(E_{g}^{\mathrm{opt, d}} - \Delta E_c)$ taken between the HOMO level of the donor and the $GW$ LUMO level of the acceptor, minus 0.3 eV, which accounts for energy conversion kinetics \cite{scharber, Thompson}. The integral in the numerator is the short circuit current $J_{sc}$ calculated using an external quantum efficiency ($EQE$) limit of 100\%, while the denominator is the integrated AM1.5 solar energy flux, which amounts to 1000 $\mathrm{W / m^2}$.
The efficiency $\eta$ is thus estimated as the product $FF \cdot V_{oc} \cdot J_{sc} $ normalized by the incident energy flux, in the limit of 100\% $EQE$
\footnote{
Ref. \cite{scharber} uses 65\% for both the $FF$ and the $EQE$. However, XSC with up to 75\% of both values have been shown recently in the literature \cite{lunt}. The alternative choice of using this 75\% limit for both $FF$ and $EQE$ would lead to a decrease in the efficiency by a factor $0.65/0.75^2 \approx 1.15$ compared to the values reported here. In addition, we note that different from Ref. \cite{scharber}, in our work $\Delta E_c > 0$ suffices to guarantee exciton dissociation at the interface, since the optical donor band gap is used instead of the electronic gap. 
}
.\\
\indent
Fig. \ref{fig2}(b) shows the efficiency of the three CBN-PCBM interfaces with type-II alignment, as a function of the CBN donor optical gap and the interface $\Delta E_c$, the latter computed as the difference between the CBN optical LUMO and the PCBM $GW$ LUMO as explained above. For the $\mathrm{C_1(BN)_7}$, $\mathrm{C_2(BN)_6}$ and $\mathrm{C_3(BN)_5}$ monolayers, the efficiency values are, respectively, 11\%, 15\% and 20\%. The striking efficiency tunability found here is achieved by changing the size of the C domain within a 1 nm range. Though such small C domains might seem challenging to achieve in practice, the immiscibility of C and BN in 2D leads to the formation of a large amount of sub-nm scale domains, to the point that the single C and BN domains cannot be resolved after the monolayer synthesis \cite{Aja}. Strategies for controlling the domain size and shape at the atomistic scale in CBN are also being actively explored \cite{sutter}.\\
\begin{figure}[!ht]
\includegraphics{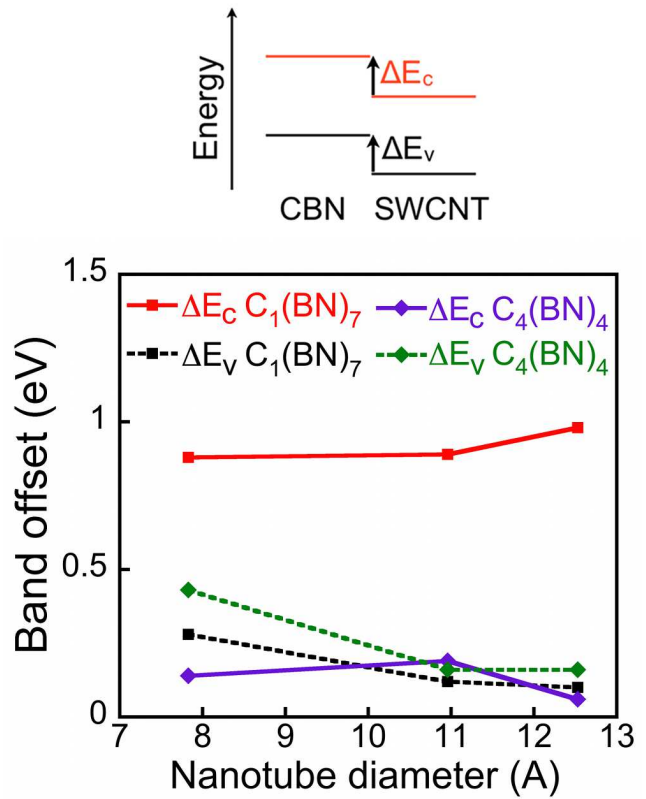}  
\caption{DFT valence ($\Delta E_v$) and conduction ($\Delta E_c$) band offsets at CBN - SWCNT interfaces, shown for combinations of the two CBN structures $\mathrm{C_1(BN)_7}$ and $\mathrm{C_4(BN)_4}$ and the three zig-zag nanotubes (10,0), (14,0) and (16,0). $\Delta E_v$ and $\Delta E_c$ are referenced, respectively, to the valence and conduction band edges of the acceptor, as shown above the plot, and are plotted as a function of nanotube diameter. Both $\Delta E_v$ (dashed lines) and $\Delta E_c$ (solid lines) are positive in all cases, indicating type-II alignment. \label{fig3}}
\end{figure}
\indent
Next, we analyze the band alignment at CBN-SWCNT and CBN bilayer interfaces. Fig. \ref{fig3} shows the DFT valence and conduction band offsets for interfaces between the two CBN systems $\mathrm{C_1(BN)_7}$ and $\mathrm{C_4(BN)_4}$ (with respectively 0.25 and 1 nm C domain size) and zig-zag s-SWCNT with three different diameters. Both $\Delta E_v$ and $\Delta E_c$ are found to be positive (with the convention shown in Fig. \ref{fig3}) for both CBN cases and regardless of the nanotube diameter, implying a type-II alignment for these interfaces. The band offsets show little variation with nanotube diameter, and $\Delta E_c$ becomes smaller for increasing C domain sizes, similar to the CBN-PCBM case. At the interface, the nanotube behaves as the acceptor, and should thus be n-doped in a real device for optimal performance. 
Though we do not verify it explicitly, the type-II alignment would be retained at the $GW$ and $GW$+BSE levels of theory. For the alignment type to be inverted, the $GW$ correction to the nanotube band gap would need to be higher than the corresponding correction to the CBN layer band gap, which are, respectively, 3.25 eV and 1.72 eV for the $\mathrm{C_1(BN)_7}$ and $\mathrm{C_4(BN)_4}$ cases \cite{ourcbn}. However, only small $GW$ corrections are predicted for s-SWCNT in this diameter range ($\approx$ 10\% of the band gap \cite{GW-SWCNT}), and such corrections certainly cannot be as large as 1.7 eV. 
Similar to the case of the CBN-PCBM interfaces, band offset values with a strong dependence on the C domain size in CBN are found, which could allow one to tune the solar cell performance by varying the structure of the CBN layer.\\ 
\begin{figure}[!t]
\includegraphics{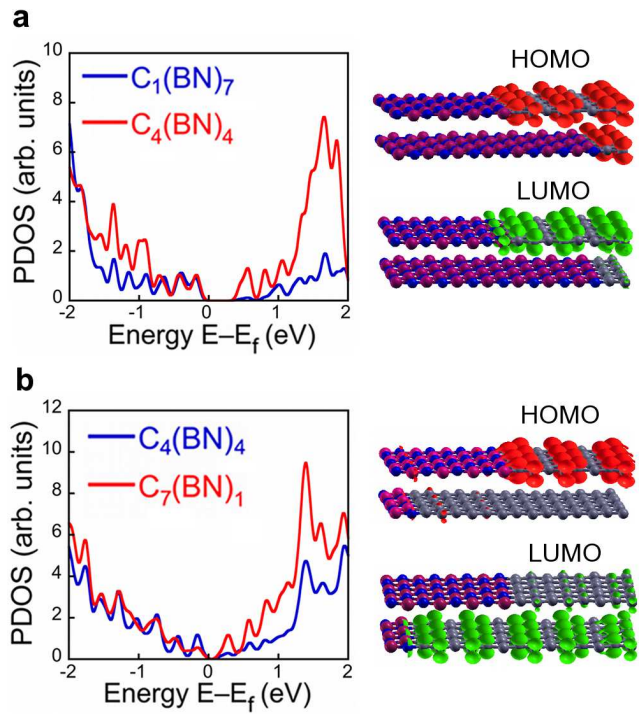}
\caption{PDOS (left) and HOMO and LUMO orbital isosurfaces (right) for CBN bilayers studied using DFT. The energies in the PDOS plot are referenced to the Fermi energy $E_f$. Two monolayer combinations are shown: (a) $\mathrm{C_1(BN)_7 / C_4(BN)_4}$ and (b) $\mathrm{C_4(BN)_4 / C_7(BN)_1}$. Type-I or type-II band alignments are predicted depending on the structure and C domain size of the CBN layers composing the bilayer.
\label{fig4}}
\end{figure}
\indent
The electronic structure of bilayer systems, in which the two composing CBN monolayers have different atomic structures, is presented in Fig. \ref{fig4}. We show the PDOS and the HOMO and LUMO orbitals of two bilayer cases among those studied in this work. In both bilayer systems, the LUMO localizes within the layer with the smaller energy gap. For the $\mathrm{C_1(BN)_7 / C_4(BN)_4}$ bilayer (Fig. \ref{fig4}(a)), we observe a complete hybridization of the valence states, as seen by the perfect overlap of the valence PDOS, causing the HOMO to delocalize to both layers. In contrast, in the $\mathrm{C_4(BN)_4 / C_7(BN)_1}$ bilayer the HOMO is found to be localized on the $\mathrm{C_4(BN)_4}$ layer, \textit{i.e.} the layer with the larger band gap. In this case, the incomplete hybridization of the valence states is seen by the slight valence band offset in the PDOS. We found an analogous behavior in the $\mathrm{C_1(BN)_7 / C_7(BN)_1}$ system.\\ 
\indent
On this basis, we predict the $\mathrm{C_1(BN)_7 / C_4(BN)_4}$ interface to be a type-I heterojunction with ohmic character (due to the absence of barriers for the transport of holes), and the $\mathrm{C_4(BN)_4 / C_7(BN)_1}$ and $\mathrm{C_1(BN)_7 / C_7(BN)_1}$ interfaces to be type-II heterojunctions, in which exciton dissociation may be possible within the bilayer. In both cases, upon photoexcitation the orbitals are predicted to dramatically change their spatial distribution (Fig. \ref{fig4}), which may lead to opportunities for engineering the flow of photoabsorbed energy. We note that the alignment types found here for the bilayer systems are retained also beyond the DFT level of theory, since the $GW$ and the $GW$+BSE corrections to the band gaps are higher for the CBN monolayer with the larger band gap \cite{ourcbn}.\\
\indent
Our results thus suggest that in principle it may be possible to fabricate a two atomic layers thick ($\approx$ 3.3 \AA) \hspace{2 pt} XSC in the form of a bilayer device, by stacking two CBN monolayers with the proper C domain structure. Alternatively, given the tunability of the CBN layer electronic structure, a Schottky junction solar cell may be formed between a monolayer of graphene and a monolayer of CBN. Considering a monolayer absorbance of approximately 2\% similar to the case of graphene, a structure with as few as $30-40$ monolayers and a thickness of approximately 10 nm could suffice to absorb most of the incident sunlight with energy above the band gap.\\
\indent
\begin{figure}[!hbt]
\includegraphics{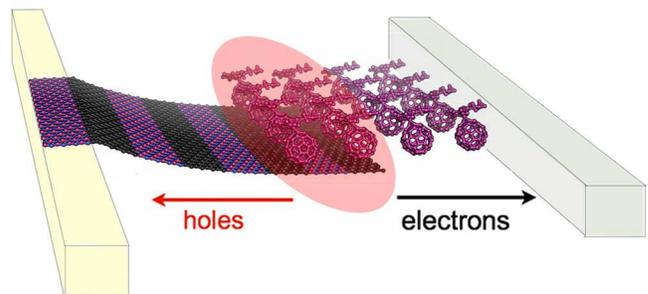}
\caption{Proof-of-concept design of a solar cell based on 2D semiconducting monolayer materials, for the CBN-PCBM  material combination. Note that the PCBM could be replaced by a second 2D monolayer, and the same architecture can be extended to other monolayer materials. The shaded oval indicates the interface where excitons are separated upon illumination (from the top in figure), and the arrows indicate the direction of carrier diffusion and extraction at the contacts, shown as thin metal fingers. \label{fig5}}
\end{figure}
In addition to the potential technological innovation resulting from the tunability of the parameters of interest in photovoltaics, XSC based on 2D monolayer materials could constitute a new platform for the experimental study of quantum transport effects (characteristic of 2D monolayers) in photovoltaic devices. For example, while hot carrier and hot exciton transfer are possible at or near an interface \cite{Ross}, in common XSC they're hindered by the diffusion through the bulk of the absorber. However, if a solar cell is made with a single monolayer, no bulk diffusion is involved and the hot carrier regime could be enabled. In addition, the impact of coherent exciton transport on XSC performance is largely unexplored in photovoltaics, due to the fact that exciton transport operates in an incoherent regime in reasonably any $\mu$m-thick bulk heterojunction solar cell fabricated using standard deposition techniques.\\ 
\indent
As a test bed for these fundamental effects, we propose in Fig. \ref{fig5} an architecture whereby XSC based on 2D monolayer materials could be fabricated and characterized, for example using ultra-fast spectroscopy measurements \cite{grancini} to ascertain the presence of hot exciton dissociation or hot carrier extraction.\\ 
\indent
We remark that while the tunability of the solar cell properties presented here originates from changes to the composition and domain structure of a single monolayer, additional tuning of physical quantities of interest in photovoltaics could be achieved by stacking sequences of different 2D semiconducting monolayers, an alternative approach we are currently investigating.\\
\indent
In summary, we present the idea of XSC based on semiconducting 2D monolayer materials with the potential to achieve $10-20$\% power conversion efficiencies in $10$ nm thick active layers, and show that combinations of CBN monolayers and PCBM or s-SWCNT are well-suited for the practical implementation of such devices. Even a photovoltaic device as thin as two atomic layers of CBN (with different C domain sizes, as shown here) or other monolayer materials with type-II band alignment holds the potential to achieve solar energy conversion at exceptionally small length and ultra-fast time scales. The unique tunability in 2D monolayer materials of the band gap, interface band alignment, exciton binding energy, optical absorption, carrier mobility, and electron-phonon coupling entails new opportunities for fundamental studies and practical implementation of excitonic solar cell devices.
\section{Acknowledgments}
M.B. acknowledges funding from Intel through the Intel Ph.D. Fellowship. We wish to thank NERSC and Teragrid for providing computational resources.
\section{Supporting Information}
Details of many-body calculations using the $GW$ and BSE methods.

\section{References}
\bibliography{references}

\begin{thebibliography}{46}
\bibitem [{Hel()}]{Heliatex}%
  \BibitemOpen
  \href@noop {} {}\bibinfo {note} {The current efficiency record for small
  molecule based solar cells is 10.7 \% from the German company Heliatex.\\ See
  heliatex.com}\BibitemShut {NoStop}%
\end{thebibliography}

\end{document}